\documentclass[useAMS,usegraphicx,usenatbib]{mn2e}
\usepackage{amsmath, amssymb}
\title[A search for the 55 MHz OH line]{A search for the 55 MHz OH
  line}

\author[V. R. Marthi and J. N. Chengalur]{Visweshwar Ram
  Marthi\thanks{E-mail: vrmarthi@ncra.tifr.res.in (VRM) \newline chengalur@ncra.tifr.res.in (JNC)} and
  Jayaram N. Chengalur$^{*}$\\
National Centre for Radio Astrophysics, Tata Instsitute of Fundamental
Research, Pune 411 007, India.}

\begin{document}
%\date{Accepted 1988 December 15. Received 1988 December 14; in original form 1988 October 11}
\pagerange{\pageref{firstpage}--\pageref{lastpage}} \pubyear{2010}
\maketitle
\label{firstpage}

\begin{abstract}
The OH molecule, found abundantly in the Milky Way, has four
transitions at the ground state rotational level($J = 3/2$) at cm
wavelengths. These are E1 transitions between the $F^+$ and $F^-$
hyperfine levels of the $\Lambda$ doublet of the
$^{2}\Pi_{\frac{3}{2}}, J=3/2$ state. There are also forbidden M1
transitions between the hyperfine levels within each of the
doublet states occuring at frequencies 53.171~MHz and
55.128~MHz. These are extremely weak and hence difficult to
detect. However there is a possibility that the level populations giving rise to these
lines are inverted under special conditions, in which case it may be
possible to detect them through their maser emission. We
describe the observational diagnostics for determining when the hyperfine
levels are inverted, and identify a region near W44 where these
conditions are satisfied. A high-velocity-resolution search for these
hyperfine OH lines using the low frequency feeds on four antennas of
the GMRT and the new GMRT Software Backend(GSB) was performed on this
target near W44. We place a 3$\sigma$ upper limit of $\sim$17.3
Jy (at 1 $km~s^{-1}$ velocity resolution) for the 55~MHz line from
this region. This corresponds to an upper limit of 3  $\times 10^8$
for the amplification of the Galactic synchrotron emission providing
the background.
\end{abstract}

\begin{keywords}
ISM: molecules, masers, techniques: spectroscopic
\end{keywords}

\section{Introduction}
The ground state rotational level (J=3/2) of the $^{2}\Pi_{\frac{3}{2}}$
ladder of the OH molecule is split by $\Lambda$-doubling into two
states, each of which is further split into two states because of the hyperfine 
interaction between the spin of the unpaired $2p$ electron of the 
O atom and the nuclear magnetic moment of the H atom. The four 
well-known 18~cm OH lines, namely the ``main lines'' at 1665~MHz \& 1667~MHz 
and the ``satellite lines'' at 1612 MHz \& 1720 MHz arise from 
transitions between these levels.  The main lines, associated with
star-forming regions, are sometimes seen strongly mased. The
1612 MHz and 1720 MHz satellite line masers are usually associated
respectively with late-type stars and
shocked regions, and sometimes seen as a mased-thermal
pair\citep{masers}. Transitions between the hyperfine states of the
same $\Lambda$ doublet state (M1) occur at 53.171 MHz($J=3/2,\ F:2^-
\rightleftarrows 1^-$) and 55.128 MHz($J=3/2,\ F:2^+ \rightleftarrows
1^+$). The thermal emission or absorption from these lines is
extremely weak given that they are highly forbidden M1 transitions with $A \leq 10^{-18}
s^{-1}$\citep{destombes}. The expected thermal line brightness temperature for
the 54 MHz lines is 1~K for an OH column density of 10$^{17}$
cm$^{-2}$\citep*{roshi}. It is not possible to detect such weak thermal
lines at 54 MHz, given the high Galactic background. Hence, the
lines have to be necessarily mased inorder to be detectable in
reasonable integration times. We discuss below the conditions for
inversion of the populations in the levels from which the 53~MHz and
55~MHz lines arise.

\begin{figure*}
\begin{center}
  \includegraphics[scale=0.6]{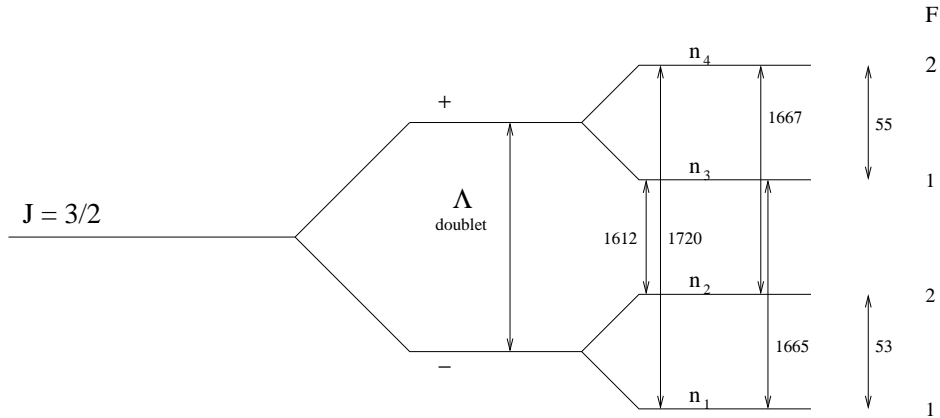}
  \caption{Ground state energy levels(J=3/2) of the
    $^2\Pi_\frac{3}{2}$ ladder of the OH molecule. n$_1$, n$_2$, n$_3$ and n$_4$
   are the hyperfine level populations. The frequencies, in MHz, of
   the transitions between these levels are also indicated.}
  \label{ladder}
\end{center}
\end{figure*}

The J=3/2 state of the $^2\Pi_{\frac{3}{2}}$ ladder is given in
Figure~\ref{ladder}.

For the 53 MHz line to be mased we require the corresponding level
populations be inverted:

\begin{equation}
\frac{n_{2}/g_{2}}{n_{1}/g_{1}} > 1 
\label{oh53}
\end{equation}

which implies

\[
\frac{n_{2}/g_{2}}{n_{3}/g_{3}}  \times
\frac{n_{3}/g_{3}}{n_{1}/g_{1}} > 1
\ \ \ and \ \ \ 
\frac{n_{2}/g_{2}}{n_{4}/g_{4}}  \times
\frac{n_{4}/g_{4}}{n_{1}/g_{1}} > 1
\]
\\
Condition \eqref{oh53} is satisfied if the populations in the levels
corresponding to the 1665 MHz and 1720 MHz lines are inverted while those in the levels
corresponding to the  1612 MHz and 1667 MHz lines are not. Observationally
if the 1665 MHz and 1720 MHz lines are mased, while the 1612 MHz and 1667 MHz
lines are not, we could expect the 53 MHz line levels to be inverted.
While the inequalities are necessary and sufficient conditions, the
requirement on the individual lines are merely sufficient, eg. 

\[
\frac{n_{2}/g_{2}}{n_{3}/g_{3}} \leq 1
\ \ \ and \ \ \
\frac{n_{3}/g_{3}}{n_{1}/g_{1}} \gg 1
\]
\\
could still give 
\[
\frac{n_{2}/g_{2}}{n_{1}/g_{1}} > 1
\]
\\
Similarly, for the 55 MHz line to be mased we require the
corresponding level populations be inverted:

\begin{equation}
\frac{n_{4}/g_{4}}{n_{3}/g_{3}} > 1
\label{oh55}
\end{equation}

which implies

\[
\frac{n_{4}/g_{4}}{n_{2}/g_{2}}  \times
\frac{n_{2}/g_{2}}{n_{3}/g_{3}} > 1
\ \ \ and \ \ \ 
\frac{n_{4}/g_{4}}{n_{1}/g_{1}}  \times
\frac{n_{1}/g_{1}}{n_{3}/g_{3}} > 1
\]
\\
Condition \eqref{oh55} is satisfied if the populations in the levels
corresponding to the 1667 MHz and 1720 MHz lines are inverted while
those in the levels corresponding to the  1612 MHz and 1665 MHz lines
are not. Observationaly if the 1667 MHz and 1720 MHz lines are mased,
while the 1612 MHz and 1665 MHz lines are not, we could expect the 55
MHz line levels to be inverted. 
\begin{table}
\begin{center}
\begin{tabular}{|c|c|c|c|c|c|} \hline
 1612 & 1665 & 1667 & 1720 & 53  & 55 \\ 
  MHz &  MHz &  MHz &  MHz & MHz & MHz \\ \hline
Not        & Inverted       & Not       & Inverted
&$\checkmark$& -- \\
inverted &                  & inverted                   & & & \\ \\
Not      & Not       & Inverted       & Inverted      & -- &
$\checkmark$ \\ 
inverted & inverted                 &                    & & & \\ \\
\hline
\end{tabular}
\end{center}
\caption{The conditions on the level populations corresponding to the
  18~cm OH lines, required for {\bf maser emission} of the 53 and/or
  55 MHz hyperfine OH lines} 
\end{table}
In summary, a region where the 1720 MHz line is mased,
but the 1612 MHz line is not, would be a promising region to look for mased 
emission from the 53/55 MHz lines. Which of these is likely to be mased
depends on whether the 1665 MHz or 1667 MHz line is mased. It is interesting
to note in this context that there are regions in our Galaxy\citep{turner}
and in external galaxies\citep*{langevelde95,chengkan} where the satellite 
lines are conjugate, viz. where their profiles are mirror images of
one another, i.e. when one is in emission the other is in absorption,
and the sum of the two profiles is consistent with noise. Regions such
as these are promising ones to search for the 53 MHz and 55 MHz OH
lines. If the hyperfine OH 53 MHz or 55 MHz lines are strongly
amplified, then the lines may be detectable. However, the
amplification factor is a matter of speculation. \citet{roshi}
attempted to detect the 53~MHz line, but could only place an upper
limit to the amplification factor. 

\section{Observations}
\subsection{The target}
For our particular observations, we have chosen the target region G34.3+0.1
($\alpha(2000) = 18^{h}\ 59^{m}\ 06^{s}.75$; $\delta(2000) = 
+01\degr\ 24\arcmin\ 39\arcsec.40$) where the 1720 MHz line is mased, 
while the 1612 MHz line is in absorption.
The main lines are both seen in emission, although it is not clear
if they are mased or not. 
The lines all occur within the LSR range of 
56-58 km s$^{-1}$\citep{turner}.  The supernova remnant W44 is only 48' away 
from G34.3+0.1 and lies well within the 10$\degr$ GMRT primary beam. In this
region the 1720 MHz line is mased, while the 1612 MHz line is seen
in absorption. The main lines are also seen in absorption over
a velocity range of ~ 10 km s$^{-1}$\citep{turner}. These lines occur over the
velocity range 43-46 km s$^{-1}$. Thus both G34.3+0.1 and W44 could give
rise to mased 53/55 MHz lines. Strictly speaking, the conditions 
described above, viz. that the 1720 MHz line is mased but the 1612 MHz 
line is not etc. should be satisfied along the same line of sight inorder 
for the 55/53 MHz lines to be mased. VLBI observations show that the OH maser emission 
generally comes from extremely compact ($\sim$ 100 mas) hot
spots\citep{hoffman}. It is rare to have VLBI observations of all the
four OH 18~cm transitions: hence it is difficult at the current time to
unambiguously identify a region where all the four transitions are
spatially co-incident. In any case, thermal emission or absorption
from OH will generally not be detected with VLBI.

Table \ref{sourcetable} lists the two candidate regions
that are situated very close to each other and hence simultaneously
observable with the $10\degr$ GMRT beam. 
\begin{table*}
\begin{center}
\begin{tabular}{c|c|c|c|c} \hline 
Galactic & 1612 & 1665 & 1667 & 1720\\
Region & MHz & MHz & MHz & MHz \\ \hline 
G34.3+0.1  & -0.80, 59.6, 3.6  & 5.70, 58.5, 2.0 & 10.65, 58.5, 1.0 &
0.61, 58.0, 2.9 \\
G34.7-0.5(W44)  & -0.82, 41.8, 9.4  & -2.00, 44.0, 10.0 & -2.70, 43.0,
10.0 & 3.28, 43.2, 2.5 \\ \hline
\end{tabular}
\end{center}
\caption{ Line parameters for two of the regions in the Galaxy that
  meet our criteria for inversion of the levels giving rise to the
  53/55 MHz lines, reproduced from \citet{turner}. The sets of three
  numbers are, for each transition, respectively the antenna
  temperature in K, the LSR velocity in $km~s^{-1}$ and the width of
  the line in $km~s^{-1}$. } 
\label{sourcetable}
\end{table*}
\subsection{Feeds, back-end and data recording}
For the observations, we used four antennas of the GMRT on which the low
frequency feeds designed and developed by the Raman Research
Institute, Bangalore, India\citep{amiri} have been installed.
These feeds have frequency coverage from 30 MHz to 90 MHz. The GMRT receiver chain was used till the baseband
unit to filter the required section of the band from 50 MHz to 58 MHz
centered at 54 MHz. Only one sideband of the receiver was used -
consequently the selected band was placed in the upper sideband(USB)
with 54 MHz falling at the centre of the sideband. The reason for
doing so will be explained below.

We used the GMRT Software Backend(GSB)\citep{gsb} for recording the raw voltage
data. The GSB is a cluster of high-performance PCs connected by
ethernet and communicating through the MPI protocol. It operates in several
modes such as raw voltage recorder, realtime interferometric correlator,
pulsar receiver, offline interferometer and beamformer, with facility
to do inbuilt bandpass filtering in the first three modes. We
exploited the high bandwidth sampler to record 8.333 MHz of the band at
the Nyquist rate. At the time the observation was carried out,
the GSB was operational for dual-sideband and one
polarization. However, by re-wiring the inputs to the GSB, we obtained
both polarizations but one sideband. This was the reason for putting
the band of interest - 50 MHz to 58 MHz - in the upper sideband of the
baseband receiver. To reduce data volume, a decimating subroutine was
added to the recording program to desample the voltage data to Nyquist
rate. It is to be noted that though the baseband filter spans 50-58
MHz, the GSB samples 50-58.333 MHz because the sampling frequency of
the GSB is 33.33 MHz.

The observations were carried out on 12 March 2009, recording the data
for about five hours. Before recording the data on the
target, a few minutes of test data was acquired with default gains and
its RMS was calculated. The gains of the samplers were then adjusted
so that the full range of the 8-bit sampler accommodated $6\sigma$,
where $\sigma$ is the RMS. A few such iterations were performed until
the gains converged. The final gain table was loaded into the samplers
and data recording was commenced. Though the sampler clocks with a
period of 33 ns, i.e. 33 Msps, every other sample was discarded to
achieve Nyquist rate and keep the data volume within limits. At the
end of five hours of observation, we had about 270 GB of data per
antenna per polarization, there being a total of four antennas with
two polarizations each. 

\section{Data processing}
Data were recorded as a contiguous time-series with a sampling period
of 66 ns(post-decimation). Each polarisation from each of the four
antennas was recorded separately on individual disks. The format of
the recorded data necessitated writing of special software to process
them. The aim was to detect, if any, very narrow spectral
lines. Since the observing frequency is centered around 54 MHz, high
velocity resolution is possible only with very high spectral
resolution. However, since we were looking for spectral features
within a limited range of LSR velocities, data were bandpass
filtered around the region of interest and desampled. We used the
Intel IPP routines in our program to construct bandpass filters of
specified pass and stop bands. The filter was designed such that an
integer number of non-overlapping filters, M, of a specified bandwidth
completely filled the observed bandwidth of 8.333 MHz. The filtered
data is then decimated by the same factor M. This operation is called
bandpass sampling. We chose M=90 for the 55 MHz OH line and 2048
channels within the passband to allow a velocity resolution of 0.25
$km~s^{-1}$ per channel. For the 53 MHz OH line, M=80 with 2048
channels, the velocity resolution obtained was 0.28 $km~s^{-1}$ per
channel. M was chosen differently to accommodate the expected line
frequencies in both cases within the central 60\% of the band.  

After filtering and desampling, approximately every 0.25 second of
data was Fourier transformed with the Intel IPP FFT routine and then
squared. The power spectrum for each block of the time series and
hence the cumulative power spectrum for each polarization was
obtained. The calculated Doppler shift of the expected line during the
observation was less than the width of one channel. The power spectra
were visually inspected for RFI. The 55 MHz passband was found to be
relatively clean and unaffected by RFI except by a very weak,
spectrally  narrow, feature at around 30 $km~s^{-1}$ LSR velocity that
manifests only in the cumulative power spectrum. Since this was away
from the region of interest, i.e. 40-70 $km~s^{-1}$, we chose to
ignore it. The cumulative spectra from all the polarizations save one,
which had a bad bandpass, were added.  Thus, effectively the four
antennas were used like four single-dish spectrometers in the
incoherent mode. The 53 MHz passband, on the contrary, was found to be
severely affected by RFI to such an extent that it had to be abandoned
altogether. 
\begin{figure*}
\begin{center}
\includegraphics[scale=0.495]{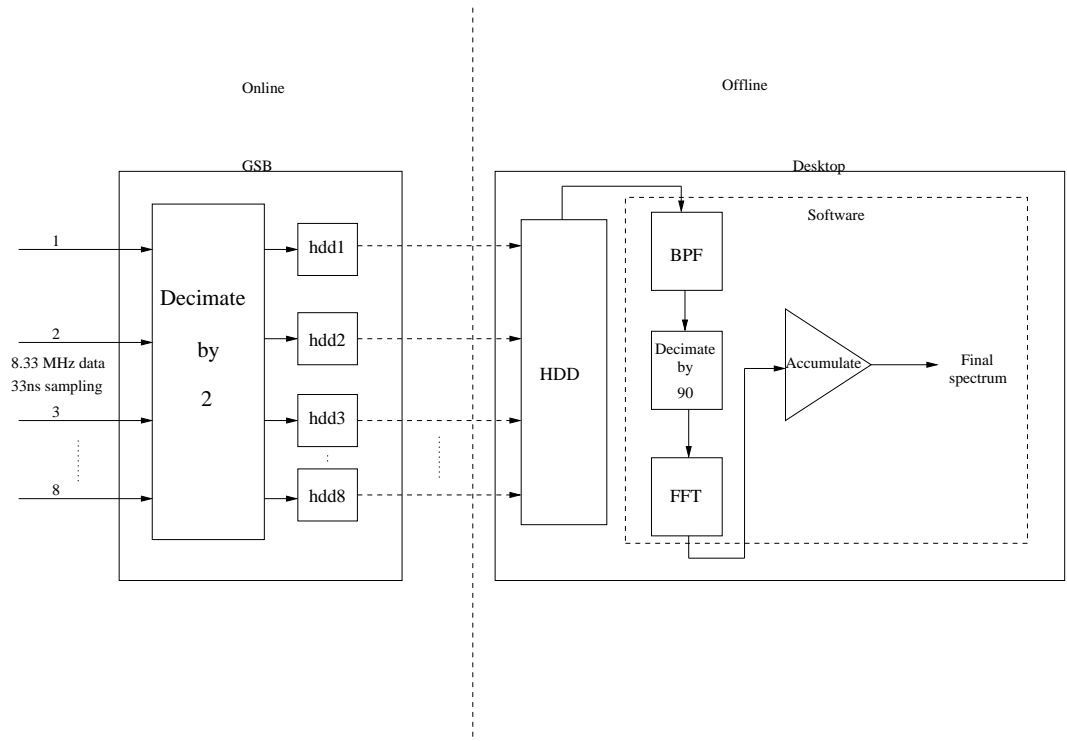}
\caption{Schematic block diagram of the data processing chain. The
  input data is sampled at the full resolution of the GSB and
  decimated by a factor of 2. The decimated voltage time series from
  the two polarizations of each antenna is recorded in individual
  disks. Offline, the data is bandpass sampled(bandpass filtering
  followed by aliased sampling) and the final cumulative power
  spectrum obtained via the FFT.}

\end{center}
\end{figure*}

\section{results and conclusions}
\begin{figure}
\begin{center}
\includegraphics[width=3.35in]{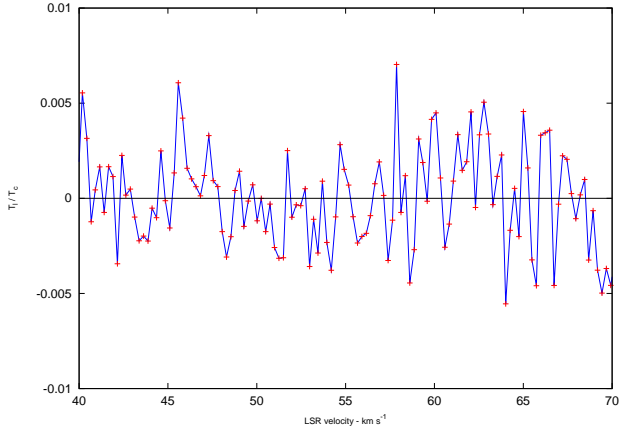}
\caption{Baseline-subtracted line-to-continuum flux ratio around
  G34.3+0.1 for the 55 MHz OH line. The velocity resolution is 0.25
$  km~s^{-1}$ per channel.} 
\label{psfit}
\end{center}
\end{figure}

\begin{figure}
\begin{center}
\includegraphics[width=3.35in]{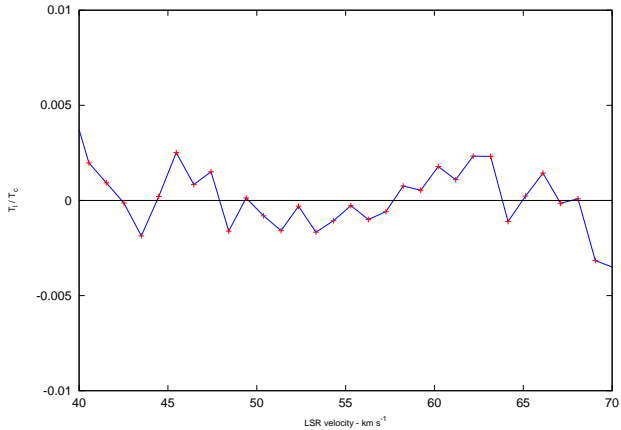}
\caption{Baseline-subtracted line-to-continuum flux ratio around
  G34.3+0.1, smoothed to a velocity resolution of 1.0 $km~s^{-1}$ per
  channel.} 
\label{psfit_smoothed}
\end{center}
\end{figure}

Figure \ref{psfit} shows the final cumulative averaged spectrum
obtained from seven out of the eight available polarizations of the four
antennas. The velocity resolution is $0.25\ km~s^{-1}$ per channel. The
ordinate is the baseline-subtracted line-to-continuum flux ratio,
after fitting for the baseline with a second order polynomial for the
passband shown in the plot. At around the expected LSR velocity of 46 $km~s^{-1}$,
where the 1720 MHz line is seen inverted towards W44 in as many as 25
hotspots\citep{claussen}, there is a weak $\sim$4$\sigma$ spectral
emission feature, whose peak is 0.006 in units of line-to-continuum
temperature ratio, (T$_l$-T$_c$)/T$_c$. The RMS optical depth
is 0.0014 units over a 0.25 $km~s^{-1}$ channel. We conclude the 55 MHz OH
line is not detected to the 4$\sigma$ limit. However, this feature
becomes more prominent(see Figure \ref{psfit_smoothed}) when the
spectrum is smoothed to $1\ km~s^{-1}$,  commensurate with the velocity
widths of the hotspots listed by \citet{claussen}. 
We assume that the system temperature is dominated by the sky temperature at these
frequencies. The sky temperature, 32100 K, is obtained by scaling the
temperature of the target region, using a power law index of -2.5,
from the 34.5 MHz survey of \citet{geetee}. For the measured aperture efficiency of
$\eta_{A}$=0.7\citep{amiri} and $A = 1590\ m^{2}$ per polarisation, we obtain, using the
equation $S A_{e\!f\!f} = kT_{l}$,  a 3$\sigma$ flux limit of 26.5
Jy for the observed line-to-continuum ratio. On smoothing to 1
$km~s^{-1}$ we obtain a flux limit of 17.3 Jy per channel.  

%\section{analysis}
The equation of radiative transfer gives
\[
T_l = T_{bg}e^{-\tau_{\nu} } + T_x(1-e^{-{\tau}_{\nu}})
\]

where $T_{bg}$ is the brightness temperature of the background source
providing the radiation and $T_x$ is the excitation temperature.

Both the background temperature $T_{bg}$ and the excitation
temperature $T_x$ contribute to the maser line brightness, but when
$T_{bg}$ far exceeds $T_x$, say $T_{bg} > 10|T_x|$, we can omit the
contribution of $T_x$ to $T_l$. The 3$\sigma$ limit to the line brightness
temperature from our observation is $T_l$ = 74 K for a 0.25 $km~s^{-1}$
channel which translates into 26.5 Jy. At a resolution of 1 $km~s^{-1}$, 
the limiting flux is 17.3 Jy. For comparison for the 53~MHz line, 
\citet{roshi} place a limit of 49~Jy at a velocity resolution of
4.6 $km~s^{-1}$.

For a maser that is $100\ mas$\citep{hoffman} in size, the beam dilution factor
for the $10 \degr$ GMRT beam is $1.3 \times 10^{1\!1}$. Given $T_{bg}$ =
32100 K, we obtain the maser amplification, omitting the contribution from
$T_x$, as  
\[
G = \frac{74 \times 1.3 \times 10^{1\!1}}{32100} \sim 3 \times 10^8
\]
providing an upper limit to the optical depth as $log(G)$ = 19.52. At a
resolution of 1.0 $km~s^{-1}$, we get an upper limit of $\tau = log(G) -
log(74/16.49) = 18.58$ given that \mbox{$T_l = 16.49$ K.}
The calculation above assumes that (some reasonable fraction of)
the Galactic synchrotron emission provides the background radiation
that is amplified by the maser. 

\section*{Acknowledgments}
The 50 MHz low frequency feeds at the GMRT have been developed
and built by the Raman Research Institute. We are grateful to
N.~Udaya~Shankar and K.~S.~Dwarakanath who led this effort.
We also wish to thank J.~Roy and S.~Kulkarni for enabling the
decimation mode in the GSB, R.~Nityananda for allotting us GMRT time
for the observations, N.~G.~Kantharia for scheduling the same and
\mbox{S.~Kudale} for help with the observations. We thank the anonymous
referee for the encouraging review comments and for the suggestions
that have helped improve the clarity of this paper. We thank the staff
of the GMRT who have made these observations possible, and to the many
generous farmers who conceded their land to the GMRT project. 
GMRT is run by the National Centre for Radio Astrophysics of the Tata
Institute of Fundamental Research.

\label{lastpage}

\end{document}